\documentclass[utf8]{frontiersSCNS} 

\usepackage{url,hyperref,lineno,microtype,subcaption}
\usepackage[onehalfspacing]{setspace}
\linenumbers

\def\firstAuthorLast{Kuruvila {et~al.}} 
\def\Authors{Ivine Kuruvila\,$^{1}$, Jan Muncke\,$^{1}$, Eghart Fischer\,$^{2}$ and Ulrich Hoppe\,$^{1,*}$}
\def\Address{$^{1}$Affiliated to the Department of Audiology, ENT-Clinic, Friedrich-Alexander-Universit\"at Erlangen-N\"urnberg (FAU), Erlangen, Germany. \\
$^{2}$Affiliated to the WS Audiology, Erlangen, Germany.}
\def\corrAuthor{Ulrich Hoppe}
\def\corrEmail{ulrich.hoppe@uk-erlangen.de}

\begin{document}
\nolinenumbers
\onecolumn
\firstpage{1}

\title[]{Extracting the Auditory Attention in a Dual~-~Speaker Scenario from EEG using a Joint CNN-LSTM Model} 

\author[\firstAuthorLast ]{\Authors} 
\address{} 
\correspondance{} 

\extraAuth{}

\maketitle

\begin{abstract}
Human brain performs remarkably well in segregating a particular speaker from interfering ones in a multispeaker scenario. We can quantitatively evaluate the segregation capability by modelling a relationship between the speech signals present in an auditory scene, and the listener's cortical signals measured using electroencephalography (EEG). This has opened up avenues to integrate neuro-feedback into hearing aids where the device can infer user's attention and enhance the attended speaker. Commonly used algorithms to infer the auditory attention are based on linear systems theory where cues such as speech envelopes are mapped on to the EEG signals. Here, we present a joint convolutional neural network (CNN) - long short-term memory (LSTM) model to infer the auditory attention. Our joint CNN-LSTM model takes the EEG signals and the spectrogram of the multiple speakers as inputs and classifies the attention to one of the speakers. We evaluated the reliability of our network using three different datasets comprising of 61 subjects, where each subject undertook a dual-speaker experiment. The three datasets analysed corresponded to speech stimuli presented in three different languages namely German, Danish and Dutch. Using the proposed joint CNN-LSTM model, we obtained a median decoding accuracy of 77.2\% at a trial duration of three seconds. Furthermore, we evaluated the amount of sparsity that the model can tolerate by means of magnitude pruning and found a tolerance of up to 50\% sparsity without substantial loss of decoding accuracy.
\newline
\end{abstract}

\section{Introduction}
Holding a conversation in presence of multiple noise sources and interfering speakers is a task that people with normal hearing carry out exceptionally well. The inherent ability to focus the auditory attention on a particular speech signal in a complex mixture is known as the cocktail party effect \citep{bib:cherry1953}. However, an automatic machine based solution to the cocktail party problem is yet to be discovered despite the intense research for more than half a century. Such a solution is highly desirable for a plethora of applications such as human-machine interface (e.g. Amazon Alexa), automatic captioning of audio/video recordings (e.g. YouTube, Netflix), advanced hearing aids etc.

In the domain of hearing aids, people with hearing loss suffer from deteriorated speech intelligibility when listening to a particular speaker in a multispeaker scenario. Hearing aids currently available in the market often do not provide sufficient amenity in such scenarios due to their inability to distinguish between the attended speaker and the ignored ones. Hence, additional information about the locus of attention is highly desirable. In visual domain, selective attention is explained in terms of visual object formation where an observer focuses on a certain object in a complex visual scene \citep{bib:feldman2003}. This was extended to auditory domain where it was suggested that phenomena such as cocktail party effect could be better understood using auditory object formation \citep{bib:shinn2008}. In other words, brain forms objects based on the multiple speakers present in an auditory scene and selects those objects belonging to a particular speaker during attentive listening (top-down or late selection). However, flexible locus of attention theory was concurrently proposed where the late selection is hypothesized to occur at low cognitive load and early selection is hypothesized to occur at high cognitive load \citep{bib:vogel2005}. This has inspired investigation into whether cortical signals could provide additional information that helps to discriminate between the attended speaker and interfering speakers. In a dual-speaker experiment, it was observed that the cortical signals measured using implanted electrodes track the salient features of the attended speaker stronger than the ignored speaker \citep{bib:mesgarani2012}. Similar results were obtained using magnetoencephalography and electroencephalography (EEG) \citep{bib:ding2012b, bib:sullivan2014}. In recent years, EEG analyses have become the commonly used methodology in attention research which is lately known as auditory attention decoding (AAD).

Both low level acoustic features (speech envelope or speech spectrogram) and high level features (phonemes or phonetics) have been used to investigate the speech tracking in cortical signals \citep{bib:aiken2008, bib:lalor2010, bib:diLiberto2015, bib:broderick2019}. State-of-the-art AAD algorithms are based on linear systems theory where acoustic features are linearly mapped on to the EEG signals. This mapping can be either in the forward direction \citep{bib:lalor2010, bib:fiedler2017, bib:kuruvila2020} or in the backward direction \citep{bib:sullivan2014, bib:biesmans2017, bib:mirkovic2015}. These algorithms have been successful in providing insights into the underlying neuroscientific processes through which brain suppresses the ignored speaker in a dual-speaker scenario. Using speech envelope as the input acoustic feature, linear algorithms could generate system response functions that characterize the auditory pathway in the forward direction. These system response functions are referred to as temporal response function (TRF) \citep{bib:lalor2010}. Analysis of the shape of TRFs has revealed that the human brain encodes the attended speaker different to that of the ignored speaker. Specifically, TRFs corresponding to the attended speaker have salient peaks around 100 ms and 200 ms which are weaker in TRFs corresponding to the ignored speaker \citep{bib:fiedler2019, bib:kuruvila2021}. Similar attention modulation effects were observed when the acoustic input was modified to using speech spectrogram or higher level features such as phonetics \citep{bib:diLiberto2015}. Likewise using backward models, the input stimulus can be reconstructed from EEG signals (stimulus reconstruction method) and a listener's attention could be inferred by comparing the reconstructed stimulus to the input stimuli \citep{bib:sullivan2014}. These findings give the possibility of integrating AAD algorithms into hearing aids which in combination with robust speech separation algorithms could greatly enhance the amenity provided to the users. 

It has been well established that the human auditory system is inherently non-linear \citep{bib:Zwicker2013} and AAD analysis based on linear systems theory addresses the issue of non-linearity to a certain extend in the preprocessing stage. For example, during speech envelope extraction. Another limitation of linear methods is the longer time delay required to classify attention \citep{bib:fuglsang2017, bib:geirnaert2019}, although there were attempts to overcome this limitation \citep{bib:miran2018, bib:kuruvila2021}. In the last few years, deep neural networks have become popular especially in the field of computer vision and natural language processing. Since neural networks have the ability to model non-linearity, they have been used to estimate the dynamic state of brain from EEG signals \citep{bib:craik2019}. Similarly in AAD paradigm, convolutional neural network (CNN) based models were proposed where the stimulus reconstruction algorithm was implemented using the CNN model to infer attention \citep{bib:deTaillez2017, bib:ciccarelli2019}. A direct classification of attention which bypasses the regression task of stimulus reconstruction, instead classifies whether the attention is to speaker 1 or speaker 2 directly was proposed in \cite{bib:ciccarelli2019, bib:Vandecappelle2021}. In a non-competing speaker experiment, classifying attention as successful vs unsuccessful or match vs mismatch was further addressed in \cite{bib:monesi2020, bib:tian2020}.

All aforementioned neural network models either did not use speech features or made use of only speech envelope as the input feature. As neural networks are data driven models, additional data/information about the speech stimuli may improve the performance of the network. In speech separation algorithms based on neural networks, spectrogram is used as the input feature to separate multiple speakers from a speech mixture \citep{bib:wang2018}. Inspired by the joint audio-visual speech separation model \citep{bib:ephrat2018}, we present a novel neural network framework that make use the speech spectrogram of multiple speakers and the EEG signals as inputs to classify the auditory attention.
 
The rest of the paper is organized as follows. In section \ref{sec:MandM}, details of the datasets that were used to train and validate the neural network are provided. In section \ref{sec:NA}, the neural network architecture is explained in detail. The results are presented in section \ref{sec:Res} and section \ref{sec:Disc} provides a discussion on the results.

\section{Materials and Methods}
\label{sec:MandM}

\subsection{Examined EEG datasets}
We evaluated the performance of our neural network model using three different EEG datasets. The first dataset was collected at our lab and it will be referred to as FAU\_Dataset \citep{bib:kuruvila2021}. The second and third datasets are publicly available and they will be referred to as DTU\_Dataset \citep{bib:fuglsang2018} and KUL\_Dataset \citep{bib:das2019} respectively.

\subsubsection{FAU\_Dataset}
This dataset comprised of EEG collected from 27 subjects who were all native German speakers. A cocktail party effect was simulated by presenting two speech stimuli simultaneously using loudspeakers and the subject was asked to attend selectively to one of the two stimuli. Speech stimuli were taken from the slowly spoken news section of the German news website \textit{www.dw.de} and were read by two male speakers. The experiment consisted of six different presentations with each presentation being approximately five minutes long making it a total of 30 minutes. EEG was collected using 21 AgCl electrodes placed over the scalp according to the 10-20 EEG format. The reference electrode was placed at the right mastoid, the ground electrode was placed at the left earlobe and the EEG signals were sampled at 2500 Hz. More details of the experiment could be found in \cite{bib:kuruvila2021}.

\subsubsection{DTU\_Dataset}
This is a publicly available dataset that was part of the work presented in \cite{bib:fuglsang2017}. The dataset consisted of 18 subjects who selectively attended to one of the two simultaneous speakers. Speech stimuli were excerpts taken from Danish audiobooks that were narrated by a male and a female speaker. The experiment consisted of 60 segments with each segment being 50 seconds long making it a total of 50 minutes. EEG were recorded using 64 electrodes and were sampled at 512 Hz. The reference electrode was chosen either as the left mastoid or as the right mastoid after visual inspection. Further details can be found in \cite{bib:fuglsang2017, bib:fuglsang2018}.

\subsubsection{KUL\_Dataset}
The final dataset that was analysed is another publicly available dataset where 16 subjects undertook selective attention experiment. Speech stimuli consisted of four Dutch stories narrated by male speakers. Each story was 12 minutes long which was further divided into two 6-minutes presentations. EEG was recorded using 64 electrodes and were sampled at 8196 Hz. The reference electrode was chosen either as TP7 or as TP8 electrode after visually inspecting the quality of the EEG signal measured at these locations. The experiment consisted of three different conditions namely HRTF, dichotic and repeated stimuli. In this work, we analysed only the dichotic condition which was 24 minutes long. Additional details about the experiment and the dataset can be found in \cite{bib:das2016, bib:das2019}. 

\newcolumntype{C}[1]{>{\centering\let\newline\\\arraybackslash\hspace{0pt}}m{#1}}

\begin{table*}[t]
\centering
\renewcommand{\arraystretch}{2}
\begin{tabular}{C{2cm} C{2cm} C{2cm} C{2cm} C{2.5cm} C{2cm}}
\hline
\hline
Name &Number of Subjects  &Duration per Subject (minutes)  &Total duration (hours)  &Experiment type  &Language\\ 
\hline
FAU\_Dataset  &27 &30 &13.5 &male + male &German\\
DTU\_Dataset  &18 &50 &15 &male + female &Danish\\
KUL\_Dataset  &16 &24 &6.4 &male + male &Dutch\\ 
\hline
\end{tabular}
\vspace{.25cm}
\caption{\textsl{Details of the EEG datasets analysed.}}
\label{table:datasets}
\end{table*}

Details of the datasets are summarized again in Table \ref{table:datasets}. A total of 34.9 hours of EEG data were examined in this work. However, the speech stimuli used were identical across subjects per dataset and they totalled 104 minutes of dual-speaker data. In all the three datasets that were analysed, the two speakers read out different stimuli. Moreover, the stimuli were presented only once to the subject in order to avoid any learning effect. For each subject, the training and the test data were split as 75\% - 25\% and we ensured that no part of the EEG or the speech used in the test data was part of the training data. The test data were further divided equally into two halves and one half was used as a validation set during the training procedure.

\begin{table}[b] 
\centering
\renewcommand{\arraystretch}{1.5}
\begin{tabular}{C{1cm} C{3.5cm} C{3cm}} 
\hline\hline
Trial duration (sec)  &EEG data (time x num\_electrodes)   &Speech data (time x freq)\\
\hline
2  & 128x10 &101x257\\
3  & 192x10 &151x257\\
4  & 256x10 &201x257\\
5  & 320x10 &251x257\\
\hline
\end{tabular}
\vspace{.25cm}
\caption{Trial duration vs dimension of the input}
\label{table:dimensions}
\end{table}

\subsection{Data Analysis}
As EEG signals analysed were collected at different sampling frequencies, they were all low pass filtered at a cut off frequency of 32 Hz and downsampled to 64 Hz sampling rate. Additionally, signals measured at only 10 electrode locations were considered for analysis and they were F7, F3, F4, F8, T7, C3, Cz, C4, T8, Pz. We analysed four different trial durations in this study namely two seconds, three seconds, four seconds and five seconds. For 2 seconds trials, an overlap of one second was applied. Thus, there were 118922 trials in total for analysis. In order to maintain the total number of trials constant, two seconds of overlap was used in case of 3 seconds trial, three seconds of overlap was used in case of 4 seconds trial and four seconds overlap was used in case of 5 seconds trial. EEG signals in each trial were further high pass filtered with a cut off frequency of 1 Hz and the filtered signals were normalized to have zero mean and unit variance at each electrode location.\\

Speech stimuli were initially low pass filtered with a cut off frequency of 8 kHz and were downsampled to a sampling rate of 16 kHz. Subsequently, they were segmented into trials with a duration of two seconds, three seconds, four seconds and five seconds at an overlap of one, two, three and four seconds respectively. The speech spectrogram for each trial was obtained by taking the absolute value of the short-time Fourier transform (STFT) coefficients. The STFT was computed using a Hann window of 32 ms duration  with a 12 ms overlap. Most of the analysis in our work was performed using 3 seconds trial and other trial durations were used only for comparison purposes. A summary of the dimensions of EEG signals and speech spectrogram after preprocessing for different trial durations is provided in Table \ref{table:dimensions}.

\section{Network Architecture}
\label{sec:NA}
\begin{figure*}[t]
\centering
\includegraphics[width=1\textwidth]{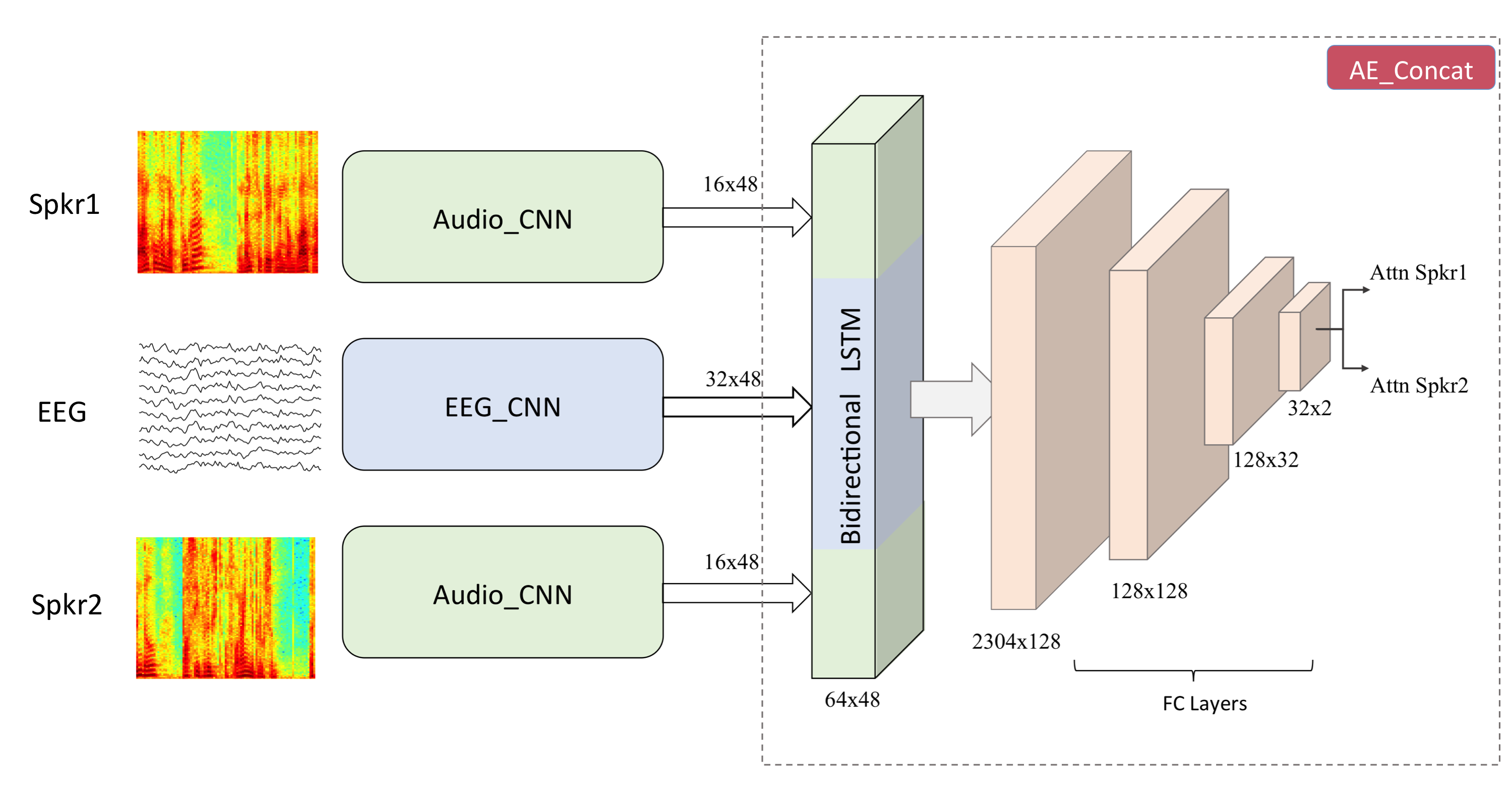}
\caption{\textsl{The architecture of the proposed joint CNN-LSTM model. Input to the audio stream is the spectrogram of speech signals and input to the EEG stream is the downsampled version of EEG signals. Number of Audio\_CNNs depends on the number of speakers present in the auditory scene (here two). From the outputs of Audio\_CNN and EEG\_CNN, speech and EEG embeddings are created which are concatenated together and passed to a BLSTM layer followed by FC layers.}}
\label{fig:DNN_Arch}
\end{figure*}

A top level view of the proposed neural network architecture is shown in Fig.\ref{fig:DNN_Arch}. It consists of three subnetworks namely EEG\_CNN, Audio\_CNN and AE\_Concat.

\subsection{EEG\_CNN}
The EEG subnetwork comprised of four different convolutional layers as shown in Table \ref{table:EEG_CNN}. The kernel size of the first layer was chosen as 24 and it corresponded to a latency of 375 ms in the time domain. A longer kernel was chosen because previous studies have shown that the TRFs corresponding to attended and unattended speakers differ around 100 ms and 200 ms \citep{bib:fiedler2019, bib:kuruvila2021}. Therefore, a latency of 375 ms could help us to extract features that modulate the attention to different speakers in a dual-speaker environment. All other layers were initialized with kernels of shorter duration as shown in Table \ref{table:EEG_CNN}. All convolutions were performed using a stride of 1x1 and after the convolutions, max pooling was used to reduce the dimensionality. To prevent overfitting on the training data and improve generalization, dropout \citep{bib:srivastava2014} and batch normalization (BN) \citep{bib:ioffe2015} were applied. Subsequently, the output was passed through a non-linear activation function which was chosen as rectified linear unit (ReLU). The dimension of the input to EEG\_CNN varied according to the length of the trial (Table \ref{table:dimensions}) but the dimension of the output was fixed at 48x32. The max pooling parameter was slightly modified for different trial durations to obtain the fixed output dimension. The first dimension (48) corresponded to the temporal axis and the second dimension (32) corresponded to the number of convolution kernels. The dimension of the output that mapped the EEG signals measured at different electrodes was reduced to one by the successive application of max pooling along the electrode axis.

\begin{table}[b] 
\centering
\renewcommand{\arraystretch}{1.5}
\begin{tabular}{C{1.5cm} C{1.75cm} C{1.5cm} C{1.45cm} C{1.45cm} C{1.6cm}} 
\hline\hline
 &Number of Kernels  &Kernel Size  &Dilation &Padding &Maxpool\\
\hline
Layer 1   &32    &24x1   &1,1    &12,0   &2,1\\
Layer 2   &32    &7x1    &2,1    &6,0    &1,2\\
Layer 3   &32    &7x5    &1,1    &3,2    &2,5\\
Layer 4   &32    &7x1    &1,1    &3,0    &1,1\\
\hline
\end{tabular}
\vspace{.25cm}
\caption{CNN parameters of the EEG subnetwork}
\label{table:EEG_CNN}
\end{table}

\subsection{Audio\_CNN}
The audio subnetwork that processed the speech spectrogram consisted of five convolution layers whose parameters are shown in Table \ref{table:Audio_CNN}. All standard procedures such as max pooling, batch normalization, dropout and ReLU activation were applied to the convolution output. Similar to the EEG\_CNN, dimension of the input to the Audio\_CNN varied according to the trial duration (Table \ref{table:dimensions}) but the dimension of the output feature map was always fixed at 48x16. As the datasets considered in this study were taken from dual-speaker experiments, the Audio\_CNN was run twice resulting in two sets of output.

\begin{table}[h] 
\centering
\renewcommand{\arraystretch}{1.5}
\begin{tabular}{C{1.5cm} C{1.75cm} C{1.5cm} C{1.45cm} C{1.45cm} C{1.6cm}} 
\hline\hline
 &Number of Kernels  &Kernel Size  &Dilation &Padding &Maxpool\\
\hline
Layer 1   &32    &1x7    &1,1    &0,3    &1,1\\
Layer 2   &32    &7x1    &1,1    &0,0    &1,4\\
Layer 3   &32    &3x5    &8,8    &0,16   &1,2\\
Layer 4   &32    &3,3    &16,16  &0,16   &1,1\\
Layer 5   &1     &1x1    &1,1    &0,0    &2,2\\
\hline
\end{tabular}
\vspace{.25cm}
\caption{CNN parameters of the Audio subnetwork}
\label{table:Audio_CNN}
\end{table}

\subsection{AE\_Concat}
The feature maps obtained from EEG\_CNN and Audio\_CNN were concatenated along the temporal axis and the dimension of the feature map after concatenation was 48x64. In this way, we ensured that half of the feature map was contributed from the EEG data and half of the feature map was contributed from the speech data. This also provides the flexibility to extend to more than two speakers such as the experiment performed in \cite{bib:schafer2018}. The concatenated feature map was passed through a bidirectional long short-term memory (BLSTM) layer \citep{bib:hochreiter1997, bib:schuster1997} which was followed by four fully connected (FC) layers. For the first three FC layers, ReLU activation was used and for the last FC layer, softmax activation was applied which helps us to classify the attention to speaker 1 or speaker 2.

The total number of EEG samples and audio samples (trials) available was 118922 and 75\% of the total available samples (89192) were used to train the network and the rest of the available samples (29730) were equally split as validation and test data. The network was trained for 80 epochs using a mini batch size of 32 samples and with a learning rate of $5*10^{-4}$. The drop out probability was set to 0.25 for the EEG\_CNN and the AE\_Concat subnetworks but it was increased to 0.4 for the Audio\_CNN subnetwork. A larger drop out probability was used for the Audio\_CNN because speech stimuli were identical across subjects for a particular dataset. Hence, when trained on data from multiple subjects, the speech data remain identical and the network may remember the speech spectrogram of the training data. The network was optimized using Adam optimizer \citep{bib:kingma2014} and the loss function used was binary cross entropy. As neural network training can result in random variations from epoch to epoch, the test accuracy was calculated as the median accuracy of the last five epochs \citep{bib:goyal2017}. The network was trained using an Nvidia Geforce RTX-2060 (6 GB) graphics card and took approximately 36 hours to complete the training. The neural network model was developed in PyTorch and the python code is available at \url{https://github.com/ivine-GIT/joint_CNN_LSTM_AAD}.

\subsection{Sparse Neural Network: Magnitude pruning}
Despite neural network learning being a sophisticated algorithm, it is still not widely used in embedded devices due to the high memory and computational power requirements. Sparse neural networks have been recently proposed to overcome these challenges and enable running these models on embedded devices \citep{bib:han2015}. In sparse networks, majority of the model parameters are zeros and zero-valued multiplications can be ignored thereby reducing the computational requirement. Similarly, only non-zero weights need to be stored on the device and for all the zero-valued weights, only their position needs to be known reducing the memory footprint. Empirical evidences have shown that neural networks tolerate high level of sparsity \citep{bib:han2015, bib:narang2017, bib:zhu2017}.

Sparse neural networks are found out by using a procedure known as network pruning. It consists of three steps. First, a large over-parameterized network is trained in order to obtain a high test accuracy as over-parameterization has stronger representation power \citep{bib:luo2017}. Second, from the trained over-parametrized network, only important weights based on certain criterion are retained and all other weights are assumed to be redundant and reinitialized to zero. Finally, the pruned network is fine-tuned by training it further using only the retained weights so as to improve the performance. Searching for the redundant weights can be based on simple criteria such as magnitude pruning \citep{bib:han2015} or based on complex algorithms such as variational dropout \citep{bib:molchanov2017} or L0 regularization \citep{bib:louizos2017}. However, it was shown that introducing sparsity using magnitude pruning could achieve comparable or better performance than complex techniques such as variational dropout or L0 regularization \citep{bib:gale2019}. Hence, we will present results based on only magnitude pruning in this work.

\begin{figure*}[t]
\centering
\includegraphics[width=.8\textwidth]{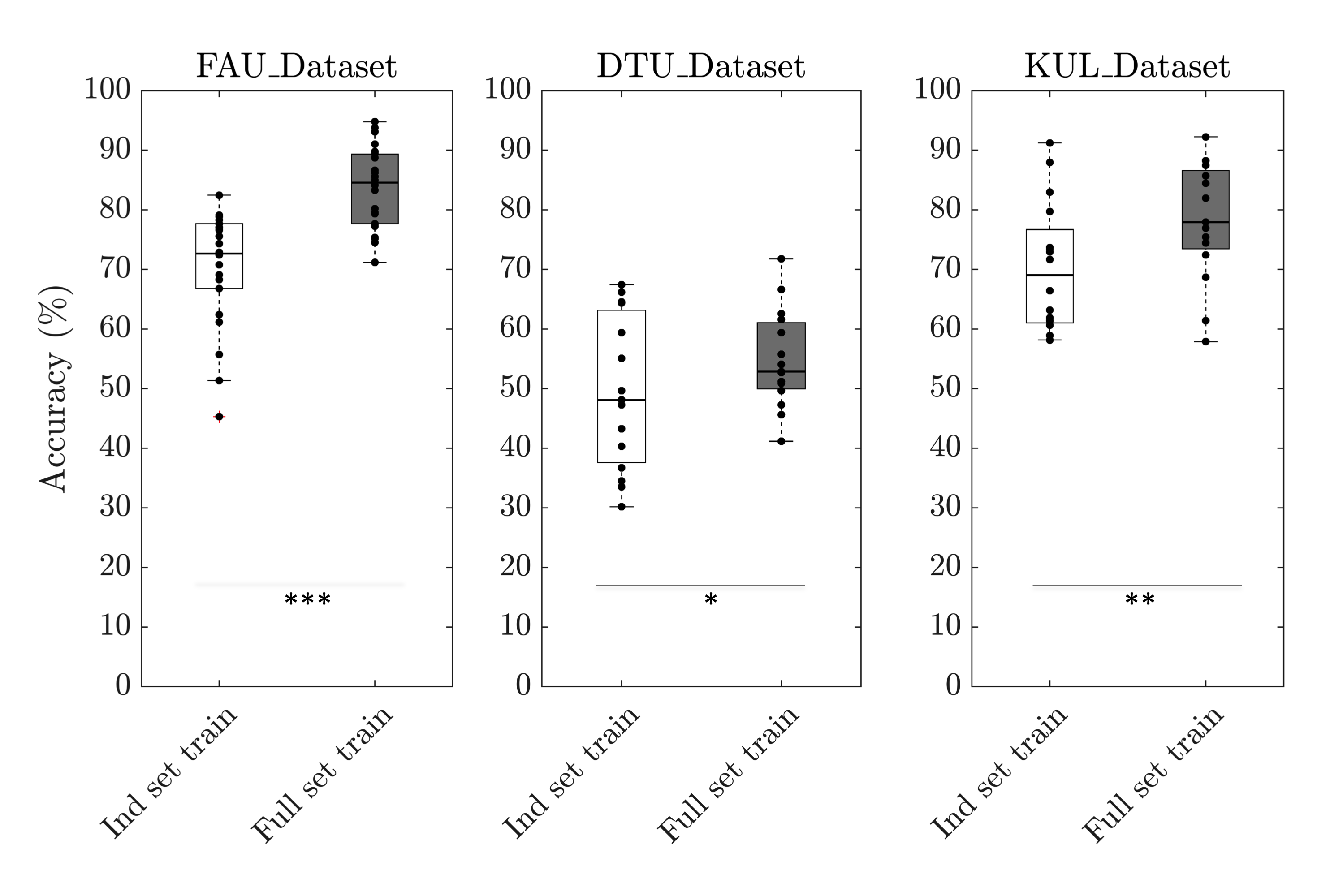}
\caption{\textsl{Boxplot depicting the decoding accuracies obtained using two different training scenarios. In the first scenario (\textit{Ind set train}), individual dataset accuracies were obtained by using training samples only from that particular dataset. For example, to calculate the test accuracy of FAU\_Dataset, training samples were taken only from FAU\_Dataset. In the second scenario (\textit{Full set train}), individual dataset accuracies were obtained using training samples from all the three datasets combined. As a result, there are more training samples in the second scenario compared to the first ($*: p<0.05,\;  **: p<0.01,\;  ***: p<0.001$ based on paired Wilcoxon signed-rank test).}}
\label{fig:acc}
\end{figure*}

\section{Results}
\label{sec:Res}
\subsection{Attention Decoding Accuracy}

\begin{figure*}[h]
\centering
\includegraphics[width=.8\textwidth]{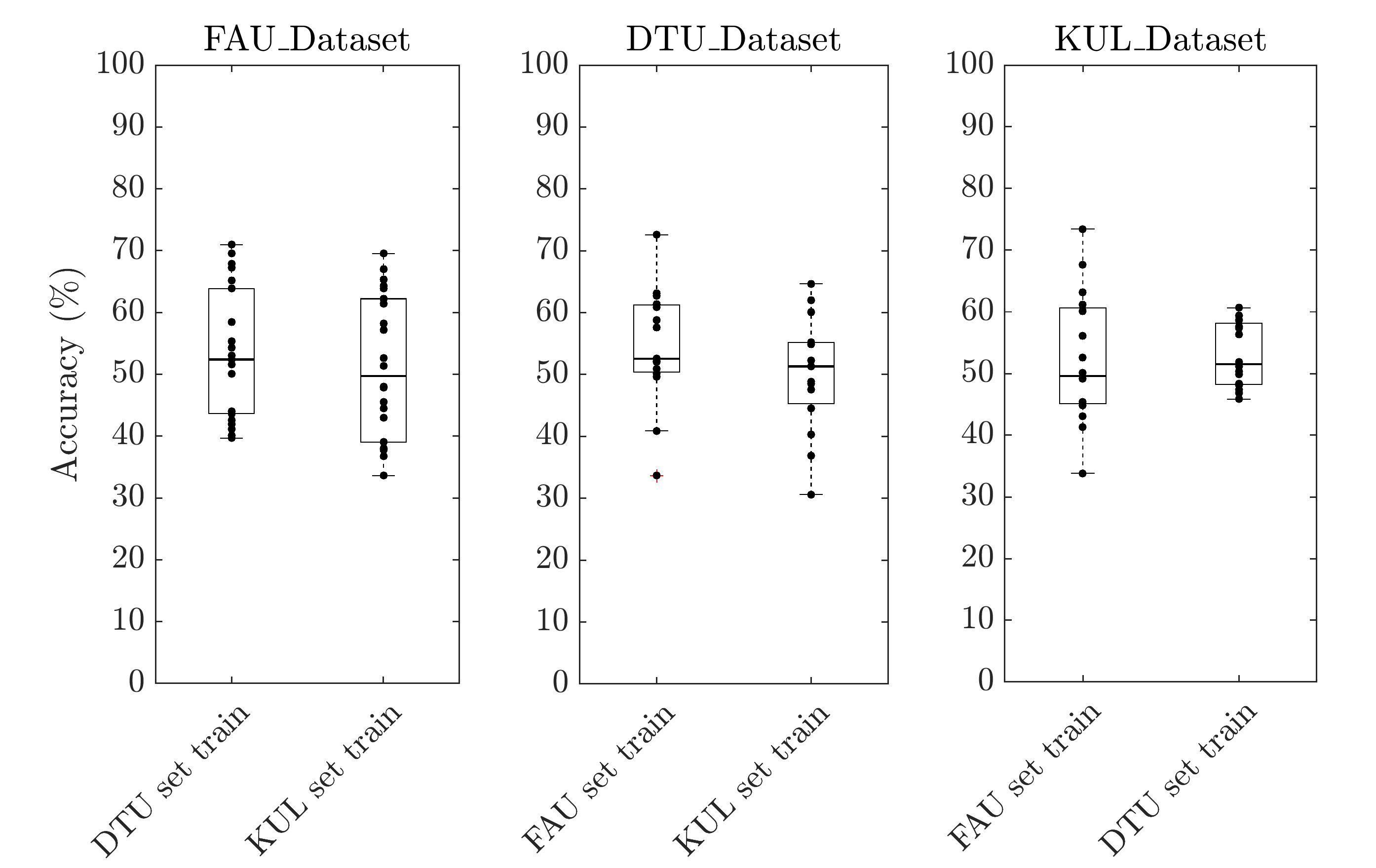}
\caption{\textsl{Boxplot showing the decoding accuracies obtained for cross-set training scenario. The accuracies obtained were all at chance level.}}
\label{fig:cross_Set_acc}
\end{figure*}

To evaluate the performance of our neural network, we trained the model under different scenarios using a trial duration of 3 seconds. In the first scenario (\textit{Ind set train}), attention decoding accuracies were calculated per individual dataset. In other words, to obtain the test accuracy of subjects belonging to FAU\_Dataset, the model was trained using training samples only from FAU\_Dataset leaving out DTU\_dataset and KUL\_Dataset. Similarly, to obtain the test accuracy for DTU\_Dataset, the model was trained using training samples only from DTU\_Dataset. The same procedure was repeated for KUL\_Dataset. The median decoding accuracy was 72.6\% for FAU\_Dataset, 48.1\% for DTU\_Dataset and 69.1\% for KUL\_Dataset (Fig. \ref{fig:acc}). In the second scenario (\textit{Full set train}), accuracies were calculated by combining training samples from all the three datasets together. The median decoding accuracies obtained in this scenario were 84.5\%, 52.9\% and 77.9\% for FAU\_Dataset, DTU\_Dataset and KUL\_Dataset respectively. The results from the second scenario showed a clear improvement over the first scenario ($p\__{FAU} < 0.001$; $p\__{DTU} < 0.05$; $p\__{KUL} < 0.01$) suggesting that the model generalizes better in the \textit{Full set train}. Furthermore, to evaluate the cross-set training performance, we trained the model using one dataset and tested it on the other two datasets. For example, the training would be performed using FAU\_Dataset and testing would be performed on both DTU and KUL datasets. The same procedure was repeated by training using the DTU dataset and the KUL dataset. The decoding accuracies obtained were all at chance level across the three cross-set training scenarios (Fig. \ref{fig:cross_Set_acc}). Consequently, all results presented further in this paper are based on \textit{Full set train}. The statistical analyses are based on paired Wilcoxon signed-rank test with sample sizes given in Table \ref{table:datasets}.
%

\begin{figure}[t]
\centering
\includegraphics[width=.75\textwidth]{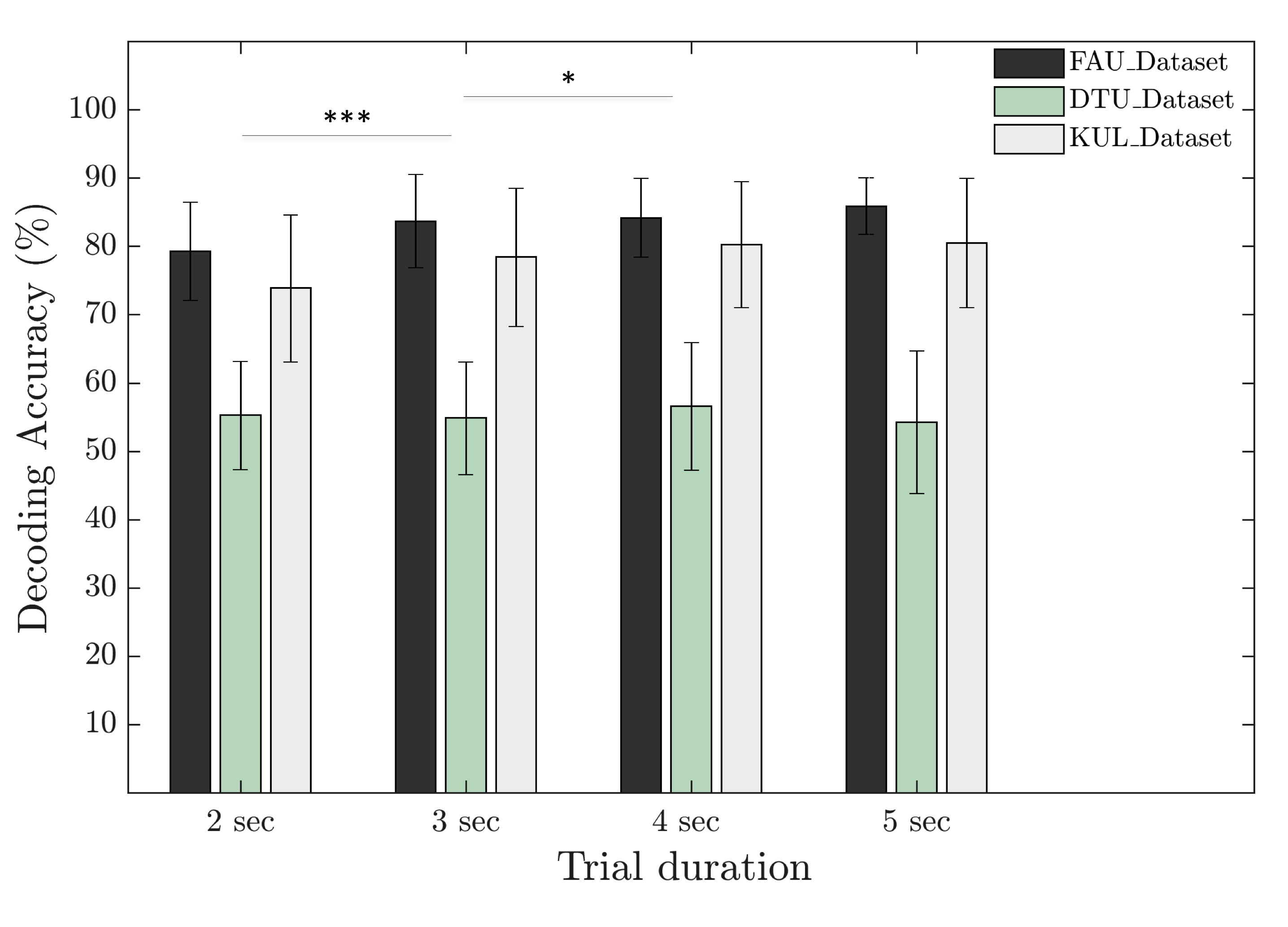}
\caption{\textsl{Comparison of the decoding accuracies calculated for different trial durations per dataset. Statistical analysis based on paired Wilcoxon signed-rank test and pooled over all subjects together from the three datasets ($*: p<0.05,\;  **: p<0.01,\;  ***: p<0.001$).}}
\label{fig:trial_dur}
\end{figure}

\subsection{Decoding Accuracy vs Trial Duration}
To analyse the effect of trial duration on the attention decoding accuracy, the model was trained using trials of length 2, 3, 4 and 5 seconds. For every trial, only one second of new data were added and the remaining data were populated by overlapping to the previous trial using a sliding window. Specifically, for 2 seconds trial, one second of overlap was used and for 3 seconds trial, two seconds of overlap was used, and so on. In this way, total number of training samples remained constant for different trial durations considered in our analysis. The mean decoding accuracy across all subjects and all datasets in case of 2 seconds trial duration was 70.9\% $\pm$ 13.2\%. The mean accuracy improved to 73.9\% $\pm$ 14.8\% when the trial duration was increased to 3 seconds ($p < 0.001, \,r=0.60$). Using a trial duration of 4 seconds, the mean accuracy obtained was 75.2\% $\pm$ 14.3\% which is a slight improvement over 3 seconds trials ($p < 0.05, \,r=0.31$). For 5 seconds trials, our neural network model resulted in a mean accuracy of 75.5\% $\pm$ 15.7\% that was statistically identical to the accuracy obtained using 4 seconds trials ($p > 0.05, \,r=0.10$). Figure \ref{fig:trial_dur} depicts the accuracy calculated for individual datasets.

\subsection{Ablation Analysis}
\begin{figure}[h]
\centering
\includegraphics[width=.75\textwidth]{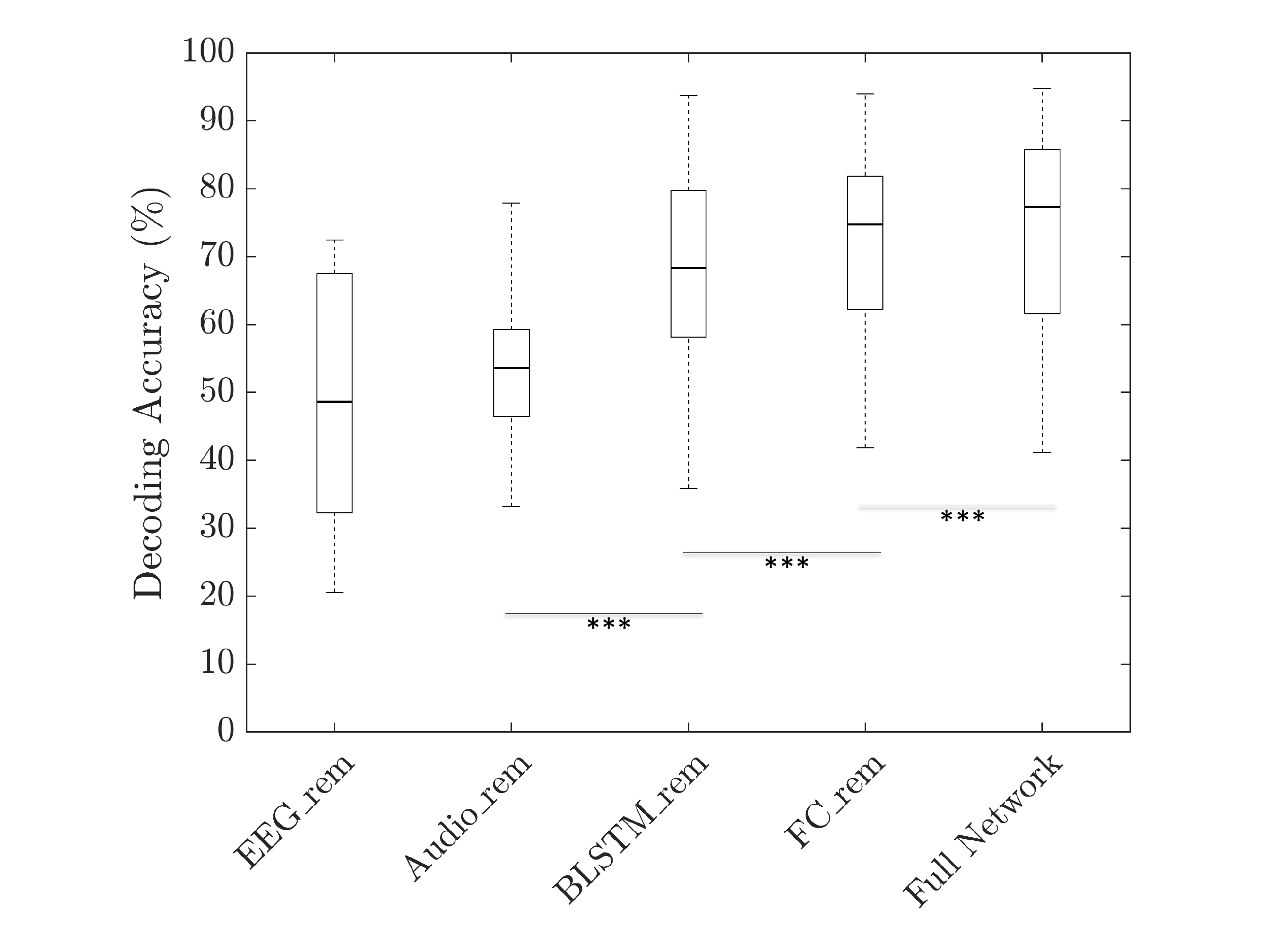}
\caption{\textsl{Boxplots showing the decoding accuracies obtained by ablating the different blocks such as FC layer or BLSTM layer. To obtain the test accuracies after ablating, the ablated network was trained from scratch in case of FC\_rem and BLSTM\_rem. However, in case of Audio\_rem and EEG\_rem, accuracies were calculated by zeroing out the corresponding input features before passing them to a fully trained network. The obtained accuracy did not demonstrate a statistically significant difference between Audio\_rem and EEG\_rem ($p > 0.05$). For all other cases, there was a significant difference ($*: p<0.05,\;  **: p<0.01,\;  ***: p<0.001$ based on paired Wilcoxon signed-rank test).}}
\label{fig:ablation}
\end{figure}

In order to gain further insights into the architecture and understand the contribution of different parts of our neural network, we performed ablation analysis using a trial duration of three seconds. To this end, we modified the neural network architecture by removing specific block such as the BLSTM layer or the FC layers one at a time and retrained the modified network. Similarly, to understand the importance of the audio input feature, decoding accuracies were calculated by zeroing out the EEG input and to understand the importance of the EEG input feature, decoding accuracies were calculated by zeroing out the audio input. As shown in Fig. \ref{fig:ablation}, the median decoding accuracy by zeroing out the EEG input was 48.6\% whereas zeroing out the audio input resulted in an accuracy of 53.6\% resulting in no significant difference ($p > 0.05$). When the network was retrained by removing the BLSTM layer only, the median decoding accuracy obtained was 68.3\% and on removing the FC layers only, median decoding accuracy was 74.7\%. Hence, the BLSTM layer contributes more towards the network learning than the FC layer ($p < 0.001$). To compare, the median decoding accuracy calculated using the full the network was 77.2\%.

\subsection{Sparse Neural Network using Magnitude Pruning}
\begin{figure}[h]
\centering
\includegraphics[width=.75\textwidth]{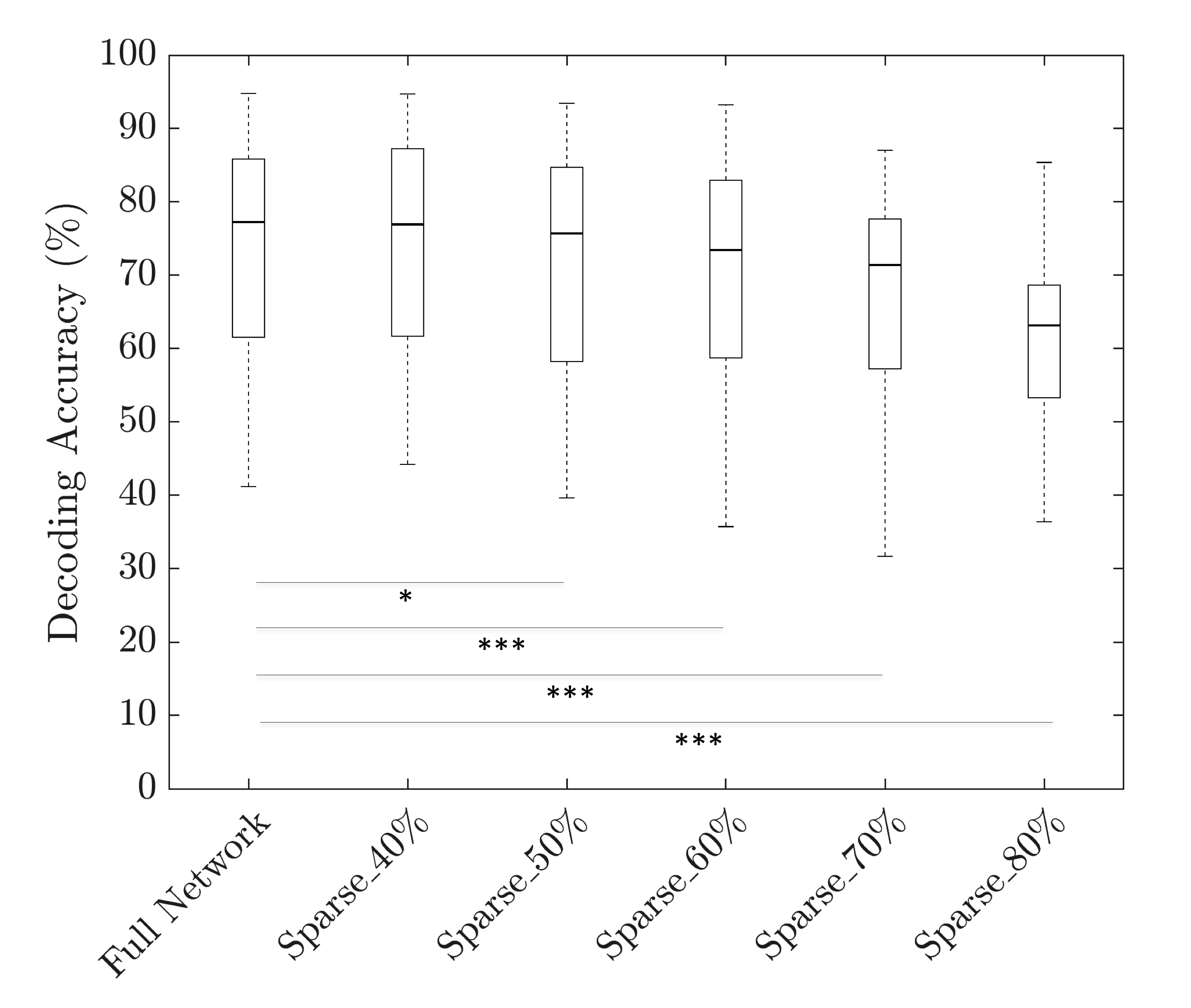}
\caption{\textsl{Plots comparing the trade off between decoding accuracies and sparsity level ($*:p<0.05,\;  **:p<0.01,\;  ***:p<0.001$ based on paired Wilcoxon signed-rank test).}}
\label{fig:pruning}
\end{figure}

To investigate the degree of sparsity that our neural network can tolerate, we pruned the model at 40\%, 50\%, 60\%, 70\% and 80\% sparsity using the three seconds trial duration. In order to fine-tune the pruned neural network, there are two options: 1) sequential or 2) one-shot. In sequential fine-tuning, weights of the trained original model are reinitialized to zero in smaller steps per epoch until the required sparsity is attained. In one-shot fine-tuning, weights of the trained original model are reinitialized to zero at one shot in the first epoch and the sparse model is further trained to improve performance. We observed that the sequential fine-tuning is less efficient than one-shot fine-tuning in terms of training time budget. Therefore,  all results presented here are based on one-shot fine-tuning. We achieved a median decoding accuracy of 76.9\% at a sparsity of 40\% which is statistically identical to the original model at 77.2\% ($p > 0.05$). When the sparsity was increased to 50\%, the median decoding accuracy decreased to 75.7\% which was lower than the original model ($p < 0.001$). Increasing the sparsity level further resulted in deterioration of decoding accuracy reaching 63.2\% at a sparsity of 80\% (Fig. \ref{fig:pruning}). Total number of learnable parameters in our model was 416741 and to find the sparse network, we pruned only the weights leaving the bias and BN parameters unchanged. 

\section{Discussion}
\label{sec:Disc}
People with hearing loss suffer from deteriorated speech intelligibility in noisy acoustic environments such as multispeaker scenarios. Increasing the audibility by means of hearing aids has not shown to provide sufficient improvement to the speech intelligibility. This is because the hearing aids are unable to estimate apriori to which speaker the user intends to listen. Hence, hearing aids amplify both the wanted signal (attended speaker) and interfering signals (ignored speakers). Recently, it has been shown that the cortical signals measured using EEG could infer the auditory attention by discriminating between the attended speaker and the ignored speaker in a dual-speaker scenario \citep{bib:sullivan2014}. Linear system analysis has been the commonly used methodology to analyse the EEG signals measured from a listener performing selective attention. However, in recent years, non-linear analyses based on neural networks have become prominent, thanks to the availability of customized hardware accelerators and associated software libraries.

In this work, we developed a joint CNN-LSTM model to infer the auditory attention of a listener in a dual-speaker environment. CNNs take the EEG signal and spectrogram of the multiple speakers as inputs and extract features through successive convolutions. These convolutions generate an intermediate embeddings of the inputs which are then given to a BLSTM layer. As LSTMs fall under the category of recurrent neural networks, they can model the temporal relationship between the EEG embedding and the multiple spectrogram embeddings. Finally, the output of the BLSTM is processed through FC layers to infer the auditory attention. The effectiveness of the proposed neural network was evaluated with the help of three different EEG datasets collected from subjects who undertook dual-speaker experiment.

There are many choices for the acoustic cues of speech signal that could be given as input to the neural network. They are speech onsets \citep{bib:howard2010}, speech envelopes \citep{bib:aiken2008}, speech spectrograms \citep{bib:pasley2012} or phonemes \citep{bib:diLiberto2015}. Due to the hierarchical processing of speech, all of the aforementioned cues could be tracked from the cortical signals measured using EEG \citep{bib:hickok2007, bib:ding2014}. Speech envelope is the most commonly used acoustic cues in the linear system analysis of EEG signal. However, we decided to use spectrogram due to its rich representational power of the corresponding speech signal and the ability of neural networks to index these multidimensional inputs efficiently.

\subsection{Attention Decoding Accuracy}
We analysed the performance of our neural network in two different training scenarios. In the first scenario, individual dataset accuracy was found out by training the network using samples taken only from that particular dataset. In the second scenario, individual dataset accuracy was found out by training using samples combined from all three datasets together. The accuracies obtained in the second scenario were higher than the first scenario by 10.8\% on average, which is in agreement with the premise of neural network learning that larger the amount of training data, the better the generalization. The decoding accuracies obtained for subjects belonging to the DTU\_Dataset were markedly lower than the other two datasets similar to the observation made in \cite{bib:geirnaert2020}. While the exact reason for the lower performance is unclear, a major difference of the DTU\_Dataset compared to the other two datasets was that the former consisted of attention to male and female speakers whereas the latter consisted of attention to only male speakers. Therefore, training with additional EEG data that consist of attention to female speakers can provide more insights into the lower performance. Additionally, we investigated the cross-set performance by training the model using one dataset and testing using the other two datasets. The accuracies obtained were all at chance level as seen in Fig. \ref{fig:cross_Set_acc}. This is not against our expectation because if the underlying training set is not representative, neural networks will not generalize. Specifically, features in the training set and the test set are different since they were recorded in different audio settings, languages and EEG devices. This further affirms the importance of having a large and diverse training set for the neural networks to function efficiently.

\subsection{Decoding Accuracy vs Trial Duration}
One of the major challenges that AAD algorithms based on linear system theory faces is the deteriorated decoding performance when the trial duration is reduced. To this end, we calculated the accuracies using our neural network for different trial durations of 2 seconds, 3 seconds, 4 seconds and 5 seconds. We observed a clear performance improvement when trial duration was increased from two to three seconds whereas for all other trial durations, accuracies did not improve substantially (Fig. \ref{fig:trial_dur}). However, increasing the trial duration will result in larger latency needed to infer the auditory attention that can adversely affect applications which require real-time operation. Hence, three seconds trial duration may be an optimal operation point as it is known from a previous study that human brain tracks the sentence phrases and phrases are normally not longer than three seconds \citep{bib:vander2019}. Similarly, our analysis made use of 10 electrodes distributed all over the scalp but future work should investigate the effect of reducing the number of electrodes. This will help in integrating algorithms based on neural networks into devices such as hearing aids. We anticipate that the current network will require modifications with additional hyperparameter tuning in order to accommodate for the reduction in number of electrodes, as the fewer is the number of electrodes, the lower is the amount of data available for training.

\subsection{Ablation Analysis}
Performing ablation analysis gives the possibility to evaluate the contribution of different inputs and modules in a neural network. To our model, when only the speech features were given as input, the median decoding accuracy was 48.6\% whereas only EEG features as input resulted in an accuracy of 53.6\% (Fig. \ref{fig:ablation}). However, statistical analysis revealed that there is no significant difference between the two. This is contrary to our anticipation because we expected the model to learn more from the EEG features than from the audio features, as the EEG signal is unique to the subject while the audio stimulus was repeated among subjects per dataset. Nevertheless, in future care must be taken to design the experiment in such a way as to incorporate diverse speech stimuli. Further analysis ablating the BLSTM layer and the FC layers revealed that the BLSTM layer was more important than the FC layers. This is probably due to the ability of the LSTM layer to model the temporal delay between speech cues and the EEG. However, we anticipate that when the training datasets become larger and more dissimilar, FC layers will become more important due to the improved representation and optimization power of dense networks \citep{bib:luo2017}.

\subsection{Sparse Neural Networks}

Although neural networks achieve state-of-the-art performances for a wide range of applications, they have large memory footprint and require extremely high computation power. Over the years, neural networks were able to extend their scope of applications was by scaling up the network size. In 1998, the CNN model (LeNet) that was successful in recognizing handwritten digits consisted of under a million parameters \citep{bib:lecun1998}, whereas AlexNet that won the ImageNet challenge in 2012 consisted of 60 million parameters \citep{bib:krizhevsky2012}. Neural networks were further scaled up to the order of 10 billion parameters and efficient methods to train these extremely large networks were presented in \cite{bib:coates2013}. While these large models are very powerful, running them on embedded devices poses huge challenges due to the large memory and computation requirements. Sparse neural networks are a novel architecture search where redundant weights are reinitialized to zero thereby reducing the computation load.

Investigation into the amount of sparsity that our neural network can tolerate revealed a tolerance of upto 50\% sparsity without substantial loss of accuracy (Fig. \ref{fig:pruning}). However, standard benchmarking on sparsity has found that deep networks such as ResNet-50 can tolerate upto 90\% sparsity \citep{bib:gale2019}. One of the potential reasons for the lower level of sparsity in our model is due to its shallow nature. I.e., our model is comprised of less than half a million learnable parameters while deep networks such as ResNet-50 is comprised of over 25 million learnable parameters. It is also interesting to note that the accuracy obtained by removing the FC layer in our ablation analysis was 74.6\% compared to the full network accuracy of 77.2\%. And the ablated network consisted of 105605 parameters which is approximately only a quarter of the total number of parameters (416741) of the original network. This shows that by careful design choices, we can reduce the network size considerably compared to an automatic sparse network search using magnitude pruning.

Sparsification of neural network has also been investigated as a neural network architecture search rather than merely as an optimization procedure. In the lottery ticket hypothesis presented in \cite{bib:frankle2018}, authors posit that, inside the structure of an over-parameterized network, there exist subnetworks (winning tickets) that when trained in isolation reaches accuracies comparable to the original network. The pre-requisite to achieve comparable accuracy is to initialize the sparse network using the original random weight initialization that was used to obtain the sparse architecture. However, it was shown that with careful choice of the learning rate, the stringent requirement on original weight initialization can be relaxed and the sparse network can be trained from scratch for any random initialization \citep{bib:liu2018}.

One of the assumptions that we have made throughout this paper is the availability of clean speech signal to obtain the spectrogram. In practice, only noisy mixtures are available and speech sources must be separated before the spectrogram can be calculated. This is an active research field and algorithms are already available based on classical signal processing such as beamforming or based on deep neural networks \cite{bib:wang2018}. Another challenge in neural network learning and in particular, its application in EEG research is the scarcity of labelled data to train the network. This limits the ability of network to generalize well to unseen EEG data. To mitigate the aforementioned limitation, data augmentation techniques are widely used in neural network training. Data augmentation is a procedure to generate synthetic dataset that spans unexplored input signal space but corresponding to the true labels \citep{bib:wen2020}. In auditory attention paradigm, linear system analyses have shown that the TRF properties differ between attended and ignored speakers \citep{bib:fiedler2019, bib:kuruvila2021}. As a result, synthetic EEG can be generated by performing a linear convolution between TRFs and the corresponding speech signal cues \citep{bib:miran2018}. The signal-to-noise ratio of the synthesized EEG can be varied by adding appropriate noise to the convolved signal. The most commonly used speech cue is the signal envelope obtained using Hilbert transform. However, more sophisticated envelope extraction methods such as the computational models simulating the auditory system could improve the quality of synthesized EEG signals \citep{bib:kates2013, bib:verhulst2018}. It must be noted that the data augmentation techniques must only be used to train the network. The validation and the testing procedure must still be performed using real datasets.

\section{Conclusion}
Integrating EEG to track the cortical signals is one of the latest proposals to enhance the quality of service provided by hearing aids to the users. EEG is envisaged to provide neuro-feedback about the user's intention thereby enabling the hearing aid to infer and enhance the attended speech signals. In the present study, we propose a joint CNN-LSTM network to classify the attended speaker in order to infer the auditory attention of a listener. The proposed neural network uses speech spectrograms and EEG signals as inputs to infer the auditory attention. Results obtained by training the network using three different EEG datasets collected from multiple subjects who undertook a dual-speaker experiment showed that our network generalizes well to different scenarios. Investigation into the importance of different constituents of our network architecture revealed that adding an LSTM layer improved the performance of the model considerably. Evaluating sparsity on the proposed joint CNN-LSTM network demonstrates that the network can tolerate upto 50\% sparsity without considerable deterioration in performance. These results could pave way to integrate algorithms based on neural networks into hearing aids that have constrained memory and computational power. 

\section*{Funding}
This work was supported by a grant from \textit{Johannes und Frieda Marohn-Stiftung, Erlangen}.

\section*{Acknowledgment}
We convey our gratitude to all participants who took part in the study and would like to thank the student Laura Rupprecht who helped us with data acquisition.

\bibliographystyle{frontiersinSCNS_ENG_HUMS}
\bibliography{literature}

\end{document}